\documentclass[10pt,a4paper]{article}
\usepackage[
left=2.00cm,
right=2.00cm,
top=4.00cm,
bottom=4.00cm
]{geometry}
\usepackage{tikz}

\usetikzlibrary{backgrounds,fit,decorations.pathreplacing,calc}

\usepackage{amsmath, amssymb}

\newcommand{\cW}{\cal W}

\newcommand{\opG}{One-Particle-Irreducible Generating Functional}

\newcommand{\gusgr}{{\rm  grand unified supergravity theory}}
\newcommand{\gust}{{SU(5) grand unified supergravity theory}}

\newcommand{\gut}{{\rm   grand unified  theory}}
\newcommand{\gutt}{{\rm SO(10) grand unified supergravity theory}}

\newcommand{\sgb}{spontaneous gauge symmetry breaking}

\newcommand{\sus}{suppressed SUSY}
\newcommand{\sS}{Suppressed SUSY}

\newcommand{\ME}{Master Equation}

\newcommand{\cM}{{\cal M}}

\newcommand{\cco}{{cosmological constant}}

\newcommand{\cG}{{\cal G}}

\newcommand{\cZ}{{\cal Z}}
\newcommand{\cA}{{\cal A}}

\newcommand{\PB}{Master Equation}

\newcommand{\bt}{\begin{tabular}{c}}
\newcommand{\et}{\end{tabular}}

\newcommand{\ebp}{\rt.\ee\be\lt.} 
\newcommand{\bmat}{\lt ( \begin{array} }
\newcommand{\emat}{  \end{array} \rt )}

\newcommand{\cP}{{\cal P}}
\newcommand{\cF}{{\cal F}}

\newcommand{\oJ}{{\ov J}}

\newcommand{\ED}{\end{document}}

\newcommand{\oY}{{\ov Y}}

\newcommand{\oC}{{\ov C}}

\newcommand{\oF}{{\ov F}}
\newcommand{\A}{{\ov A}}

\newcommand{\cD}{{\cal D}}

\renewcommand{\a}{\alpha}	
\renewcommand{\b}{\beta}

\renewcommand{\d}{\delta}

\newcommand{\h}{\eta}

\renewcommand{\k}{\kappa}
\newcommand{\lam}{\lambda}
\newcommand{\m}{\mu}
\newcommand{\n}{\nu}

\newcommand{\s}{\sigma}

\renewcommand{\c}{\chi}

\newcommand{\w}{\omega}

\newcommand{\Lam}{\Lambda}

\renewcommand{\P}{\Pi}	
\renewcommand{\S}{\Sigma}

\newcommand{\la}{\label}
\newcommand{\ci}{\cite}

\newcommand{\ds}{\documentstyle}	
\newcommand{\fr}{\frac}

\newcommand{\pa}{\partial}
\newcommand{\ov}{\overline}
\newcommand{\be}{\begin{equation}}
\newcommand{\ee}{\end{equation}}
\newcommand{\ba}{\begin{array}} 
\newcommand{\ea}{\end{array}}
\newcommand{\bea}{\begin{eqnarray}}
\newcommand{\eea}{\end{eqnarray}}
\newcommand{\ra}{\rightarrow}
\newcommand{\Ra}{\Rightarrow}

\newcommand{\lt}{\left}
\newcommand{\rt}{\right}

\newcommand{\ben}{\begin{enumerate}}
\newcommand{\een}{\end{enumerate}}
\newcommand{\bitem}{\begin{itemize}}

\newcounter{orange} 
\newcounter{apple} 
\newcounter{grape} 
\addtocounter{grape}{1}

\newcommand{\articlenumber}{5780SuppSUSYCosmo.tex}

\usepackage{color} 

\def\hourandminute{\count255=\time\divide\count255 by 60
\xdef\hour{\number\count255}
\multiply\count255 by -60\advance\count255 by\time
\hour:\ifnum\count255<10 0\fi\the\count255}

\begin{document}


\renewcommand{\thefootnote}{\fnsymbol{footnote}}

 \begin{center}

{ \huge  Suppressed SUSY  and the Cosmological Constant\\[.5cm]} 
\vspace*{.1in}
{ John A. Dixon\footnote{jadixg@gmail.com and  john.dixon@ucalgary.ca}
 \\ University of Calgary,\\
Calgary, Alberta, Canada}
\\[.5cm] 

\end{center}

\normalsize

 \begin{center}    Abstract  \end{center}

Rigid SUSY with gauge symmetry has a simple form for the Higgs  potential.  It is the positive semi--definite sum of the squares of the auxiliary fields.  Setting the potential to zero yields a set of equations for the Vacuum Expectation Values (`VEVs') of the scalar fields.  These VEVs yield  zero vacuum energy for the Higgs potential, even after gauge symmetry breaking (`GSB'). Nothing like this happens in  theories without SUSY.  However, there are four problems that make this  initially promising feature look quite useless and ineffective:  (1) SUSY has too many Higgs Fields.  (2) SUSY predicts supermultiplets.
(3) Spontaneous  breaking of SUSY looks contrived and it is unsuccessful anyway (because of the sum rules and the huge  auxiliary VEVs). 
(4) There are negative terms in the supergravity version of the Higgs potential.  
These problems are  normally  considered to be an inevitable consequence of SUSY.  However, using a technique we call suppression of fields (or `flipping'), we can remove all four problems, without disturbing the basic algebra of SUSY. 

We start with a conventional Grand Unified Supergravity Theory (`GUST'), and its BRST \ME.  The BRST Master Equation treats fields and Zinn sources more or less like coordinates and momenta in the Classical Poisson Bracket, and that suggests exchanging the fields and the Zinn sources, as in a canonical transfomation.  We call this flipping.  We can choose a special Higgs superpotential of the form $  W_H =
 e^{- \k^2  I_2 }   \sqrt{\fr{I_3}{\k^3} },   
$ where $I_n$ are polynomial gauge Invariants of order n, and $\k$ is the Planck length.  Then we apply suppression to remove all unwanted superpartner fields and to reduce the Higgs fields so that  the Higgs superpotential becomes real: $  W_H =\ov W_H$. 
The resulting new `suppressed' GUST has a simple quadratic Higgs potential, and so it has a naturally zero cosmological constant  after GSB (at tree level).  It also has a Master Equation, simply derived from the original GUST, that preserves its symmetry, and its unitarity.

\normalsize
\small
\normalsize
\large
\normalsize
\small
\large
  \begin{center} { Introduction} \end{center}

 \refstepcounter{orange}
{ \theorange}.\;
{\bf  An intriguing and venerable  Puzzle:}    Why is the \cco  \ci{ccnonzero} so near to zero?   The Higgs scalar has now been discovered \ci{higgsfound}, and this encourages us to believe that gauge symmetry is indeed spontaneously broken, in some sense, in the Standard Model \ci{stdmodelspontbk}.  It is natural to inquire how the Higgs potential relates to that.  But that brings us back to the question of the vacuum energy, and that clearly relates to the \cco.  

 \refstepcounter{orange}
{ \theorange}.\;
{\bf Rigid SUSY  offers a tantalizing reason} why, and how, the vacuum energy density could be exactly zero both before and after gauge symmetry breaking.  But   the four problems listed above in the {Abstract} for this paper  show that this nice form for the scalar potential  is  useless in conventional SUSY actions \ci{superspace,WB,ferrarabook,west,Weinberg3,Weinberg:1988cp}.

\refstepcounter{orange}
{ \theorange}.\;
{\bf  Can SUSY be the answer to the \cco\ problem, after all?} 
In this paper, we show how to construct a   { suppressed} Grand Unified Supergravity Theory \ci{ross,GUT} (`suppressed GUST')  for which  the  \cco\ is naturally zero at tree level, even after gauge symmetry breaking.  All suppressed  GUST actions have the nice quadratic semi-positive form for the scalar potential, and, at least at tree level, they do not suffer from the four problems.

\refstepcounter{orange}
{ \theorange}.\;
\la{mastereqpara3}
        {\bf  When SUSY became popular,} the community accepted the notions of `effective theories', the MSSM, soft breaking and the invisible sector \ci{Weinberg3,xerxes,haber,buchmueller}, basically as a plausible way to proceed, and to bypass the four problems in the {Abstract} for this paper.     
For related reasons, the anthropic argument  has become quite popular, in the absence of any reason  for a naturally small \cco\ \ci{anthrop}. 
 These discouraged any more work on the \ME, because integrating out heavy fields removes the symmetry that gives rise to the \ME\ 
   \cite{Dixon:2018dzx}.

    \refstepcounter{orange}
{ \theorange}.\;
\la{mastereqpara4}
 {\bf   The development of the Standard Model} needed Faddeev-Popov ghosts and an understanding of spontaneous gauge symmetry breaking,  Slavnov-Taylor identities and renormalization theory.  These took lots of work by many people over many years \ci{taylor}.  The BRST Master Equation \ci{taylor,Zinnarticle,poissonbrak, Becchi:1975nq, zinnbook,
becchiarticles1,becchiarticles2,BV,Weinberg2,babble} encapsulates many of these important features succinctly, and it ensures that the theory yields a unitary S matrix.    {\em The Master Equation is the most general quantum mechanical solution for the problem of unphysical degrees of freedom, whenever they appear} \ci{Becchi}. The general view has been that the  \ME\ is merely of technical interest.  In this paper we show that it is also  dynamical, because we can use its power to change the particle content of the theory, while maintaining unitarity.

 \refstepcounter{orange}
{ \theorange}.\;
\la{orgofpaper2}
{\bf  Preliminary Remarks:}
  The BRST Master Equation approach in Appendix \ref{brstme} below introduces `sources' like $ j_{\m},\cdots $,  of various kinds, in terms like $\int d^4 x \lt ( j_{\m} V^{\m} + \cdots \rt )$. We add these to the action  to get a functional $\cZ[j,\cdots]$.  These sources are tools to generate Green's functions like the self-energy etc.  One takes functional derivatives with respect to them, and then sets the sources to zero,  
and that leaves functions like
\be
\lt. \fr{\d^2 \cZ}{\d j^{\m_1}(x_1)\d j^{\m_2}(x_2)}\rt |_{ j \ra 0} = \lt < 0 | T\lt \{ V_{\m}(x_1)  V_{\n}(x_2)\rt \} | 0\rt >
\ee 
To get unitarity in the quantum field theory, one needs to worry about the problem that there are unphysical degrees of freedom in fields like the vector field $V_{\m}$. To deal with this, for example in non-Abelian gauge theories,  one adds the `Zinn source' term $\int d^4 x \S^{a \m} \d V_{\m}^a$ where 
$\w^a$ is the Faddeev-Popov ghost. The BRS transformation here is $ \d V^{\m} = D^{ab}_{\m}\w^b$. Taking functional derivatives like $\fr{\cdots \d  \cZ}{\d \S^{a \m_1}(x_1)\cdots }$ implements the BRS transformations.

\refstepcounter{orange}
{ \theorange}.\;
{\bf The Master Equation is very much like the Classical Poisson Bracket, with fields as coordinates and Zinn sources as momenta:}  The Master Equation turns out to be  identically true for all values of the variables, and to all loop orders. This incorporates   renormalization theory, and the Slavnov-Taylor-Ward-BRST identities (Appendix \ref{brstme}).   
The physics (particles, S matrix etc.)  lies solely in the physical fields, and not in  the Zinn sources, ghosts or antighosts.  However, the Zinn sources and the quantized variables appear in a very symmetrical way in equations (\ref{brstmas}) and  (\ref{mesimple}).

   \refstepcounter{orange}
{ \theorange}.\;
\la{temptingtocanonical}
 {\bf   These features make it tempting to see if the roles of the fields and the Zinn sources can be exchanged in some way, as if one were perfoming  a canonical transformation in a Poisson Bracket (See Appendix \ref{flippinginv}):}   In this paper we  show that this formalism leads to nice ways to build theories that are not simply based on quantizing actions and incorporating the initial symmetry--we quantize them in a more interesting way by keeping all of the symmetry, but dividing the quantization between Zinn sources and the old-style physical fields, while also taking care of the problem of unphysical degrees of freedom. The resulting physical theory inherits the symmetry from  its origin, but not in a simple way.  Moreover, some fields are more adapted to this `flip' than others.  The flipped (formerly quantized) fields get replaced by a quantized, but non-propagating, antighost field, as will be explained below.

\refstepcounter{orange}
{ \theorange}.\;
\la{orgofpaper1}
{\bf   The present paper shows that  SUSY can be used in a new and simpler way, using suppression (`flipping').} This generates a nice simple Higgs potential, no superpartners, very few parameters, and a naturally zero \cco\ (at tree level). So \sus\ looks like  progress, and its predictions are much simpler and vastly different from those of conventional SUSY phenomenology. 
      However supergravity is a non-renormalizable theory.  So it is not at all clear that this success is sustainable beyond tree level. 
The best hope for a sensible extension of a suppressed GUST beyond the tree level probably lies in superstring theory, and for that we need some sort of BRST \ME\ for the string  \ci{westbranefields,West:2012vka} and superstring \ci{Green:2012oqa,Green:2012pqa}.

 \refstepcounter{orange}
{ \theorange}.\;
\la{orgofpaper}
{\bf  Highlights:}     The main body of this paper is a short  exposition of how the four problems are dealt with by \sus.    The crucial parts are the introduction to the \ME\ in paragraph \ref{mastereqpara1}, followed by an introduction to the suppression of the scalar field in paragraph \ref{suppsusystep2}, and the reason  that gauge, gravitino and graviton fields cannot be suppressed in paragraph \ref{gaugecannotbesuppressed}.  Then in paragraph  \ref{fermionsokforsuppression},  a special issue for the suppression of spin $\fr{1}{2}$ fermions is discussed. In paragraph \ref{quadsumforgravpot}, we show how to remove the negative part of the scalar potential in supergravity, which ensures that the \cco\ is zero at tree level, both before and after \sgb. The Conclusion and Summary are in paragraph 
\ref{parconclus}. Appendix \ref{flippinginv} contains a simple formal argument explaining how and why suppression (flipping) works.

  \begin{center} {  The \ME} \end{center}

      \refstepcounter{orange}
{ \theorange}.\;
\la{mastereqpara1}
{\bf The Master Equation for a GUST:} A summary of the derivation is in Appendix \ref{brstme}.  Sums over indices and variable names are suppressed.
\be
\cM=\int d^4 x \lt \{
\fr{\d \cG}{\d \cF}
\fr{\d \cG}{\d \cZ_{\cF}}
+
\fr{\d \cG}{\d G}
\fr{\d \cG}{\d \cZ_{G }}
+
\fr{\d \cG}{\d {\cP}}
\fr{\d \cG}{\d \cZ_{{\cP}}}
\rt \}
=0
\la{brstmas}
\ee
In the above, $\cG$ is the \opG. It can be calculated using Feynman graphs and the loop expansion is $\cG=\cG_0+ \hbar \cG_1 + \hbar^2 \cG_2+ \cdots$.  The zero loop approximation is the action  ${\cA}=\cG_0$. We assume that  we have the vectors, gravitino and graviton all gauge fixed, and that, to start with, there is no spontaneous breaking or VEVs.  The derivation of the BRST \ME\ is quite simple, formally, using the path integral.   The variables in the action and the \ME\ are as follows:
\ben
\item
\la{physfields}
$\cF$ are the physical fields of the theory.   They include the Higgs scalars, the quarks and leptons, the gauge vector bosons,  and the graviton; and their superpartners, which are the Higgsinos, the squarks and sleptons, the gauginos and the gravitino. We assume that all auxiliaries have been integrated as described in paragraph \ref{integofaux} and Appendix \ref{Ftermsinteg}.
$\cZ_{\cF}$ are the   sources for the variations of the physical fields 
$\cF$.  All fields $\cF$  have ghost charge 0, and all sources $\cZ_{\cF}$ have ghost charge -1.

\item
$G$ are the Faddeev--Popov ghosts of 
the theory.  There are three kinds, the ghost for the gauge interactions, the ghost for the gravitino, and the ghost for the graviton. 
$\cZ_{G }$ are the   sources for the variations of the three kinds of ghosts $G$. All ghosts $G$  have ghost charge 1, and all sources $\cZ_{G}$ have ghost charge -2.
\item
${\cP}$ are the antighosts of the theory.  There are three kinds that are matched to, and form propagators with, the ghosts.    
\la{antighostfields}
$\cZ_{\cP}$ are the   sources for the variations of the antighosts
${\cP}$.   All antighosts $G$  have ghost charge -1, and all sources $\cZ_{\cP}$ have ghost charge 0.
\een
The physical fields, ghosts and antighosts are all quantized variables, and the sources for their variations, which we call Zinn sources, are all classical unquantized variables.

  \refstepcounter{orange}
{ \theorange}.\;
{\bf  Redundant Higgs multiplets in SUSY theories:} The first of the four problems in the {Abstract} is that there are an excessive number of  Higgs multiplets in these theories \ci{superspace,WB,ferrarabook,west,Weinberg3}. This happens because SUSY demands that the superpotential be made from chiral superfields, which are necessarily complex, even for gauge representations that are otherwise real and irreducible. 
      So our starting point is to try to `reduce these multiplets to the minimum' (see paragraph \ref{quadsumforgravpot}) to get the gauge symmetry breaking to work nicely. This can be done with { suppression}, and it is a major step down the road to deal with all the four problems. Obviously this will, in a sense, break SUSY explicitly, but we will see that there is still a great deal of structure left when we { suppress} fields, rather than remove them entirely.

  \begin{center} { Suppression of Scalar Fields} \end{center}

  \refstepcounter{orange}
{ \theorange}.\;
 {\bf Consider  the Action for the simple rigid case of SUSY coupled to gauge theory.} We include the Zinn sources for the gauge bosons, gauginos and chiral matter here. 
 This allows us to discuss \sus\ using the form of this action and \ME.    It should be remembered that the \ME\ for a SUSY theory necessarily must include supergravity  

  \refstepcounter{orange}
{ \theorange}.\;
\la{theAterms}
{\bf The $A$ terms of the Rigid SUSY Action:} 
We will first discuss the way that \sus\ works for the simplest case, which is the case of a scalar field in a chiral multiplet in the rigid SUSY theory.  
Here we pick out  the terms that include  the scalar fields $A^i$   in a rigid SUSY theory with chiral scalar fields $A^i$,  chiral spinors $\chi_{\a}^a$, gauge fields $V_{\m}^a$, gauginos $\lam_{\a}^a$, a gauge invariant superpotential function $W[A]$,  a conjugate (Zinn) source $\Lam_i$ coupled to the BRS variation $\d A^i$, a conjugate (Zinn) source $Y_{i\a}$ coupled to the BRS variation $\d \c^{\a i}$, and  a conjugate (Zinn) source $J^a_{ \a}$ coupled to the BRS variation $\d \lam^{a \a }$: 
\[
{\cA}_{\rm A}
=\int d^4 x \lt \{
 \lt (\cD_{\m}  {A}\rt )^i \lt (\ov{\cD}^{\m}   {\A}\rt )_i
+ i T^j_{ai} \lam^{a}_{\a} \chi^{i\a}   {\A}_j 
- i T^j_{ai} {\ov \lam}^{a}_{\dot \a} {\ov \chi}_j^{ \dot\a}   {A}^i 
\rt.\]
\be
\lt.
+ Y_i^{\a} {\cD}^i_{j\a\dot \b} A^j\oC^{\dot \b}
+
\Lambda_i 
i T^i_{aj}  A^j \w^a 
+ {\rm c.c.} 
\rt \}
\ee

\be
{\cA}_{\rm A\;  from \; F }= 
-
\int d^4 x \lt \{
\lt (
Y_i^{\a} C_{\a} + \fr{\pa {W}}{\pa A^i} 
\rt )
\lt (
\oY^{i \dot\a}  \oC_{ \dot\a} + \fr{\pa \ov{W}}{\pa \A_i}
\rt )
\rt \}
\la{AfromF}
\ee

\[
{\cA}_{\rm A\;from \; D }= 
-
\int d^4 x 
\lt (
iJ^{a \a} C_{\a} - 
i\oJ^{a \dot \a} \oC_{\dot \a} + 
T^{aj}_i \A_j A^i
\rt )
\]
\be
\lt (
iJ^{a \b} C_{\b} - 
i\oJ^{a \dot \b} \oC_{\dot \b} + 
T^{ak}_l \A_k A^l
\rt )
\la{AfromD}
\ee
In the above, $\cD^i_{j\m} A^j$ is the gauge covariant derivative of the scalar $A$, $C_{\a}$ are supersymmetry spinor ghosts, $\w^a$ are gauge 
ghosts and $T^a$ are Hermitian matrices for the gauge group representation. 
The terms  (\ref{AfromF}) and (\ref{AfromD}) arise from integrating the auxiliary fields $F$ and $D$, as is explained further in Paragraph \ref{integofaux}. This notation is used and explained further particularly in \cite{Dixon:2018dzx} and also in 
  \cite{Dixon:2015cya,Dixon:2016chk,Dixon:2016vkw} and 
  \cite{Dixon:2017eej,Dixon:2019vhb}.

\refstepcounter{orange}
{ \theorange}.\;
\la{integofaux}
{\bf Integration of the auxiliaries:}   It is standard practice to integrate these out of the theory, so that the physics is easier to see. We do not need to introduce sources for the auxiliaries and we do not need to introduce Zinn sources for their variations either, but we do use the fact that they participate in the nilpotence of the BRST $\d$.    See Appendix \ref{Ftermsinteg} for further explanation. 

\refstepcounter{orange}
{ \theorange}.\;
\la{thexiterms}
{\bf The $\Lambda$ terms:}   
 We can also pick out the terms that involve the  conjugate Zinn source $\Lambda$.  They are: 
\be
{\cA}_{\rm \Lambda}= 
\int d^4 x \lt \{
\Lambda_i \lt (
 C^{\a} \chi^i_{\a} +i T^i_{aj}  A^j \w^a 
\rt )
+{\rm c.c.}
\rt \}
\la{chiralzinnxiterms}
\ee

 \refstepcounter{orange}
{ \theorange}.\;
\la{suppsusystep7} \
{\bf Pieces of the \ME:}
Paragraphs (\ref{theAterms})
and (\ref{thexiterms})
 are the parts of the action that are needed to form the lowest order terms for the following piece of the \ME:
\be
\int d^4 x \lt \{ + \cdots +
\fr{\d \cG}{\d A^i}
\fr{\d \cG}{\d \Lambda_{i} }+
\fr{\d \cG}{\d \A_i}
\fr{\d \cG}{\d {\ov \Lambda}^{i} }+  +\cdots
\rt \}
\la{alam}\ee
These are terms of the type $\int d^4 x \lt \{
\fr{\d \cG}{\d \cF}
\fr{\d \cG}{\d \cZ_{\cF }}
\rt \}
$
in the notation of paragraph \ref{mastereqpara1}.

\refstepcounter{orange}
{ \theorange}.\;
\la{suppsusystep2}
{\bf The Suppression of some of the physical fields $A$:} The trick of suppressed SUSY is simply to choose to treat some set of the fields $A^i$ { not as quantized fields, but as unquantized variables}.  That is the same as treating them as Zinn sources. 
In short, we { dequantize} some of the quantized fields $A^i$.  We must also  perform a second, compensating, trick to balance this decision.  We can keep the \ME\ in the same form by  { quantizing} the unquantized conjugate sources $\Lambda_i$.  	Quantizing $\Lambda_i$  defines  new quantum fields, which  have the quantum numbers of antighosts.   
   Dequantizing fields and quantizing Zinn sources does not change their dimension, or their ghost charge, or any other quantum number, or the terms in the action that contain them.  But it does make the field into an unquantized Zinn source, and the Zinn source into a quantized antighost field.  { We sometimes call this suppression operation `flipping' the field and its Zinn source.}

  \refstepcounter{orange}
{ \theorange}.\;
\la{rederiviveme}
{ \bf   We are guaranteed that the Master Equation is still true.} 
 Quantizing the Zinn source $\Lambda$ allows us to rederive the \ME\ for the new theory.  This is shown in Appendix \ref{flippinginv} using the concept of `Flipping Invariance' which is defined there.
As noted and discussed below in paragraph 
 \ref{formofactiondoesnotchange}, the form of the  action does not change. 
 Note that this has nothing particular to do with supersymmetry--it is also true for Yang-Mills theory with scalars and no SUSY \cite{Dixon:2018dzx}. Indeed it works for any BRST nilpotent theory.

 \refstepcounter{orange}
{ \theorange}.\;
{\bf It is worth emphasizing   that the new \sus\ action appears to be identical to the old action.}  
\la{formofactiondoesnotchange}
The only difference is that certain { suppressed fields are  not quantized} in the suppressed  action, and their Zinn sources { get quantized} in compensation.  { This difference is not noticeable in the usual notation for these theories.}   This shows clearly that the { original SUSY theory is still highly relevant}. That is why we still get the nice form for the scalar potential in the suppressed  theory. 
 
 \refstepcounter{orange}
{ \theorange}.\;  {\bf Notation for the suppressed  theory:}
To actually use the suppressed  theory, we certainly want to revise the notation to show which fields are quantized and which are not, because our Feynman rules and spectrum and VEVs and gauge fixing all depend on that. When we do that, the two actions will no longer look the same; they will merely look closely related. Then, when we solve for the VEVs, in the new suppressed action, and implement gauge symmetry breaking, field shifts, and the related ghost and gauge fixing action, the differences will yield very different physics in the Feynman expansion. To see the new physics easily, one can simply set all the Zinn sources in the new suppresssed action to zero--what remains is the physical theory, made exclusively from quantized variables.

\refstepcounter{orange}
{ \theorange}.\;
{\bf  Unitarity for the Suppressed Theory:}
Consider the terms in (\ref{chiralzinnxiterms}). 
We can show that the new $\Lambda$ antighost term does not  have a propagator, because it has no quadratic term to yield a propagator. This is important because it indicates that the new suppressed theory is still unitary, as discussed in Appendix \ref{propdestroysunitarity}. 
  The ghost $C^{\a}$ does not propagate unless we are in the full supergravity theory.  Then we have more terms.  For the supergravity details see pages 385-391 of \ci{freepro} and also \ci { west,West:2012vka,pran,Buchbinder:1998qv}. 
The ghost $C_{\a}$ propagates into an antighost $\ov{E}_{\dot\a}$ There are plenty more terms in supergravity, and this $\ov{E}_{\dot\a}$  is one of them.  But it is clear that there is no way to form a counterterm like 
\be
\kappa^4 \int d^4 x \lt \{
\Lambda_i \Box
 {\ov \Lambda}^i
\rt \}
\la{nopetritermforxi}\ee
for example.  Note that this is obvious, because the term (\ref{nopetritermforxi}) has ghost charge -2, and so it cannot appear as a counterterm in the theory, because the theory conserves ghost charge.

\refstepcounter{orange}
{ \theorange}.\
\la{propforlambdaparagraph}
{\bf  Unitarity for the  suppressed theory with GSB:}
When the VEV $\lt < A^i \rt > = m v^i$ arises
in the new suppressed GUST, we get a term:
\be
\Lambda_i i T^i_{aj}  A^j \w^a 
\ra 
\Lambda_i i T^i_{aj} m v^j  \w^a 
\la{chiralzinnxitermsvev}
\ee
Is this a quadratic term that acts like a propagator?  See Appendix 
\ref{propforlambdaappendix}
 for a discussion.  We find that in fact either the $\Lambda$ is a Zinn source, or the $A$ is a Zinn source in all such terms. So there is still no propagation of this antighost in the suppressed theory with
spontaneously broken gauge symmetry.

  \refstepcounter{orange}
{ \theorange}.\;
{\bf The non-renormalization theorems (`NRTs'):} The NRTs were found early and they were a major motivation in  effective theories like the 
MSSM \ci{Weinberg3,superspace,WB,ferrarabook}. Clearly the NRTs are not  present when some fields are suppressed, because the NRTs depend on the presence of supermultiplets.  There was  talk of the NRTs `being useful for the hierarchy problem'.  But the NRTs were really not that useful, because they never did explain where the hierarchy came from--the NRTs only provided an argument that renormalization did not change the hierarchy, if it was there to start with \ci{xerxes}.

\refstepcounter{orange}
{ \theorange}.\;
\la{gaugecannotbesuppressed}
 {\bf   Suppression does not work with gauge fields,  gravitinos or gravitons,} because the homogeneous derivative term in their variation, like $\d V_{\m}^a = \pa_{\m} \w^a +\cdots$,  gives rise to a term $\S^{\m a}\d V_{\m}^a=  \S^{\m a}\lt (\pa_{\m} \w^a+ \cdots\rt )$.  When the Zinn source $\S^{\m a} $ is converted to a new antighost, this term clearly gives rise to a propagator.   The gravitino and the graviton have the same issue, so they also cannot be removed in this way. {  Suppression of  gauge fields, gravitinos, or gravitons { does introduce new propagators.}}
As we discuss further in Appendix \ref{propdestroysunitarity}, it looks wrong to try to use suppression for any fields with homogeneous gauge type symmetries.   But that is fine, because there is no need to do so. However, it is crucial to see if we can use suppression for spinor fields.

  \begin{center} { Suppression of Spinor Fields} \end{center}

  \refstepcounter{orange}
{ \theorange}.\;
{\bf Fermions:} Similar reasoning shows that we can suppress spin $\fr{1}{2}$ fermions in these theories, the same way that we did for the scalars. 
There are some interesting differences however.

\refstepcounter{orange}
{ \theorange}.\;
\la{chiandlamfromrigid}
{ \bf The $\chi$ and $\lam$  Terms:} 
Here are the spin $\fr{1}{2}$ field terms in the rigid SUSY action. We have the new Zinn source $\S^{a \m}$ for the vector variation $\d V^a_{\m} $ here, and the antighost $\h^a$:
\[
{\cA}_{\rm  Fermions}= 
\int d^4 x \lt \{
 \lam^{a \a} \cD^{ab}_{\a \dot \b} {\ov \lam}^{b \dot \b}
+  \chi^{i \a} \cD^j_{i \a \dot \b} {\ov \chi}^{\dot \b}_j
\rt. \]\[
 -  T^j_{ai} \lam^{a}_{\a} \chi^{i\a} \A_j 
-  T^j_{ai} {\ov \lam}^{a}_{\dot \a} {\ov \chi}_j^{ \dot\a} A^i 
\]\[
+  \chi^{i \a}  \chi^{j}_{ \a} \fr{\pa^2 {W}}{\pa A^i\pa A^j}
+ \S^{a \m} 
\lt (
 \lam^{a \a} \s_{\m \a \dot \b} \oC^{\dot \b}
+ C^{\a} \s_{\m \a \dot \b} {\ov \lam}^{a \dot \b}
\rt )
\]\[+
J_a^{\a} 
 f^{abc}   \lam^b_{\a} \w^a 
+
\Lambda_i  C^{\a} \chi^i_{\a}
+
Y_i^{\a}  T^i_{aj} \chi^i_{\a}\w^a 
\]
\be
\lt.  -
 \h^a \pa^{\m} \lt (  \lam^{a \a} \s_{\m \a \dot \b} \oC^{\dot \b}
+ C^{\a} \s_{\m \a \dot \b} {\ov \lam}^{a \dot \b}
\rt )  
\rt \}  
\ee

\refstepcounter{orange}
{ \theorange}.\;
\la{YandJfromrigid}
{\bf The $Y$ and $J $ Zinn terms for the fermions:}   
We can also pick out  the terms that involve the $Y$ and $J $ Zinn terms in a rigid SUSY theory with chiral fields and gauge fields. We have the   Gauge Strength $F^{a}_{ \m \n}$  here, 
and (\ref{AfromF}) and 
(\ref{AfromD}) are also needed again here:
\be
{\cA}_{\rm J }= 
\int d^4 x \lt \{
J_a^{\a} \lt ( F^{a}_{ \m\n} \s^{\m\n}_{\a \b} C^{\b}
+ f^{abc}   \lam^b_{\a} \w^a 
\rt )
+{\rm c.c.}
\rt \}
\la{Jgaugezinn}
\ee
\be
{\cA}_{\rm Y }= 
\int d^4 x \lt \{
Y_i^{\a} 
\lt (
  T^i_{aj} \chi^i_{\a}\w^a 
+ \cD^i_{j\a\dot \b} A^j\oC^{\dot \b}
\rt )
+{\rm c.c.} 
\rt \}
\la{Ychiralzinn}
\ee

 \refstepcounter{orange}
{ \theorange}.\;
\la{suppsusystep6}
{\bf Pieces of the \ME:} Paragraphs \ref{chiandlamfromrigid}
and \ref{YandJfromrigid}
 are the parts of the action that are needed to form the lowest order terms for the following piece of the \ME:
\be
\int d^4 x \lt \{ + \cdots +
\fr{\d \cG}{\d \chi^{\a i}}
\fr{\d \cG}{\d Y_{\a i} }+
\fr{\d \cG}{\d \lambda^{a \a}}
\fr{\d \cG}{\d J^a_{\a } }+
{\rm c.c.}
+ \cdots 
\rt \}
\la{formnochangespin}
\ee
The comments of paragraph \ref{suppsusystep7} are applicable here, {\em mutatis mutandis}.
 The flip of \sus\ (see paragraphs \ref{suppsusystep2} and \ref{suppsusystep2fermi}) takes these terms into a term of the form $\int d^4 x \lt \{
\fr{\d \cG}{\d \cP}
\fr{\d \cG}{\d \cZ_{\cP }}
\rt \}
$, but this does not make any change in the form of (\ref{alam}) or 
(\ref{formnochangespin}), unless we change notation as discussed in paragraph \ref{formofactiondoesnotchange}.

\refstepcounter{orange}
{ \theorange}.\;
\la{suppsusystep2fermi}
{\bf The Suppression of the physical fields $\chi$ and $\lambda$:} We do exactly the same trick for the fermions that we did for the scalars above in paragraph \ref{suppsusystep2}.
We `dequantize'   some set of the physical fields   $\chi$ and $\lambda$ and `quantize' their conjugate Zinn sources $Y$ and $J$.  Then we rederive the \ME, just as we did in paragraph \ref{rederiviveme}.

    \refstepcounter{orange}
{ \theorange}.\;
\la{fermionsokforsuppression}
{\bf Unitarity for suppression  of fermions:} Can we can  show  that there are no propagators for the new antighosts $Y$ and $J$, so that unitarity can survive into the suppressed theory after GSB?   That depends on whether there are any terms of the forms:
\be
{\cP}_{\rm Quad}= 
-
\int d^4 x \lt \{
Y_i^{\a} C_{\a} m^2 f^i  
+
J^{a \b} C_{\b} m^2 d^a  
+ c.c. \rt \}
\ee
for some constant nonzero A-independent vector $f^i$ or 
 $d^a$. 
We can see that these do not arise from  (\ref{AfromF}) and 
(\ref{AfromD}), assuming that there is no VEV for the auxiliaries:
\be
 \lt < \ov F_i \rt >=  \lt < \fr{\pa {W}}{\pa A^i} \rt >_{|_{ A^j \ra \lt < A^j \rt> }}  =m^2 \ov f_i=0
\la{vevpaW}
\ \ee
 \be
 \lt < D^a  \rt >= \lt <T^{ai}_j A^j\A_i \rt >_{|_{ A^j \ra \lt < A^j \rt> }}  =m^2 d^a=0
\la{vevD}
 \ee
This is crucial, because if it were not true, then the relevant antighosts would propagate.  But these expressions are zero, because   no spontaneous breaking of SUSY is present, which means that the VEVs of the auxiliary fields F in (\ref{vevpaW}) and D in (\ref{vevD}) are all zero.

   \begin{center} { Supermultiplets and spontaneous breaking of SUSY are no longer necessary} \end{center}
   
 \refstepcounter{orange}
{ \theorange}.\;
 {  \bf The new theory:}
 So the reduction using suppressed SUSY can change the scalar content, and the spinor content,  but not the gauge bosons or the gravitino or the graviton. In this way we will start with a SUSY theory and whittle it down until it looks like a grand unified theory without SUSY.  However, there is a major restriction on the scalar potential, and a massive  gravitino, and a zero cosmological constant, and the same unchanged action, except  with lots of unquantized Zinn sources that used to be physical scalars or spinors, and lots of  quantized non-propagating antighosts that used to be Zinn sources for physical scalars or spinors.  There is also the same \ME\ and the same resulting cohomology operator. But the flipping has changed the Feynman rules.

   \refstepcounter{orange}
{ \theorange}.\;
{\bf The second of the four problems is that the   theory contains supermultiplets even after gauge symmetry breaking. } These  consist of particles with the same mass, and different spins.  Our world does not  contain any obvious supermultiplets. 
So we will remove them using \sus.   The quarks and leptons are physical particles.  So we flip the spin 0 squarks and  sleptons using the methods above.  The gauge and Higgs particles are also physical particles. So we   flip the spin $\fr{1}{2}$ Higgsinos and gauginos.

    \refstepcounter{orange}
{ \theorange}.\;
{\bf The third of the four problems is spontaneous breaking of SUSY and its unpleasant consequences.} 
Once we have removed the unwanted scalar and spinor superpartners, there is no need to worry about spontaneous breaking of SUSY. One might think that SUSY is simply gone.  But that is not so, because the form of the Master Equation and the action have not changed at all, and the form of the scalar potential is exactly the same as it was.  But the interpretation is different, because a carefully selected set of the quantized fields and the Zinn sources  have changed their quantization character.

   \begin{center} {  A special form for the Superpotential} \end{center}
   
   \refstepcounter{orange}
{ \theorange}.\;
\la{scalarpotinsugra}
{\bf The fourth of the four problems has to do with the scalar potential of supergravity:}  Let us look at the scalar potential for a supergravity theory in 3 +1 dimensions.  This has the form \ci{freepro}:
    \[
 V = e^{\lt (\k^2 \sum_i z^i \ov z_i\rt )}
 \lt \{ \sum_j \lt | \fr{\pa W}{\pa z^j}+ \k^2 {\ov z}_j W \rt |^2
- 3 \k^2 \lt | W \rt |^2 \rt \}
\]\be
+ \fr{1}{2} \sum_{\a} \lt | D_{\a}\rt |^2
\la{complexV1}  
\ee
	Here $W(z)$ is the superpotential,  $D_{\a}(z, \ov z)$ are the auxiliary $D$ terms for the Yang-Mills theory, and $z^i$ are the chiral scalar fields in the action, with complex conjugates $\ov z_i$. The constant $ \fr{1}{\k} = \fr{M_{\rm Planck} }{\sqrt{8 \pi}}= 2.4 \times 10^{18}\; {\rm GeV}= M_P$  is the `reduced' Planck mass\footnote{Quoting p 574 of \ci{freepro}, we note that $\k^2 = \sqrt{8 \pi G}$ and $M_{\rm Planck} = G^{-\fr{1}{2}}= 1.2 \times 10^{19}$  GeV. }.  This appears in the action as $\int d^4 x \sqrt{-g} V$
so a VEV for V would be a \cco.

\refstepcounter{orange}
{ \theorange}.\;
{\bf Use of the   `Field Counting  Operator':}  
The expression (\ref{complexV1}) can also be written:
\[
V
=
e^{\lt (\k^2 \sum_i z^i \ov z_i\rt )} \lt \{
 \sum_j \lt | \fr{\pa W}{\pa z^j} \rt |^2
\rt. \]
\be
\lt. + 
 \k^2  
  \lt [
 N    
- 3     
+\k^2 
\sum_j z^j
{\ov z}_j 
 \rt ]
\lt | W 
 \rt |^2
\rt \}
+ \fr{1}{2} \sum_{\a} \lt ( D_{\a}\rt )^2
\la{complexV2}
\ee
where we define the field Counting Operator:
\be
N=
 \sum_j 
 \lt (
  z^j  \fr{ \pa}{\pa z^j}
 +
 \ov z_j  \fr{\pa}{\pa \ov z_j}  
\rt ) 
\la{defnN}
\ee

\refstepcounter{orange}
{ \theorange}.\;
\la{quadsumforgravpot}
{\bf Reduction to the Quadratic Sum:}  To reduce (\ref{complexV2}) to the nice form we need the superpotential $W$ to satisfy:
\be
  \lt [
 N    
- 3     
+\k^2 
\sum_j z^j
{\ov z}_j 
 \rt ]
\lt | W 
 \rt |^2=0
\la{countingequation1}
\ee
To achieve this, we can start with a superpotential of the form:
 \be
  W =
 e^{\fr{-1}{4}\k^2  I_2 } 
   \sqrt{\fr{I_3}{\k^3} }  
\la{formofhiggssuperpot4}
    \ee
where $I_2= 
\sum_j z^j z^j$ is the quadratic invariant expresssion in the Higgs fields, and  $I_3$ is a general cubic invariant expression in the Higgs fields. This satisfies
 \be
  \lt [
 N    
- \fr{3}{2}  
+\fr{ \k^2 }{2}
I_2
 \rt ]
  W
  =0
\la{halfportion}    \ee
Then we  suppress scalar fields so that  $W= \ov W$, and $I_2= \ov I_2 = \sum_j z^j
{\ov z}_j $. Then (\ref{halfportion}) implies (\ref{countingequation1}). It turns out that also $D^{\a}\equiv 0$ (at least in \cite{Dixon:2018dzx,Dixon:2017eej,Dixon:2019vhb}), so  (\ref{complexV2}), reduced to just physical scalar Higgs fields, becomes:
\be
V
=
e^{\lt (\k^2 I_2\rt )} \lt \{
 \sum_j \lt ( \fr{\pa W}{\pa z^j} \rt )^2
\rt \}
\la{complexV3}
\ee
This form (\ref{complexV3}) is sufficiently close to the form  of the rigid potential in (\ref{AfromF}), so that the same considerations are relevant to both cases.

   \begin{center} { Final Steps, Philosophy and Conclusion} \end{center}

\refstepcounter{orange}
{ \theorange}.\;
\la{scalarpotinsugra2}
{\bf Spontaneous Gauge Symmetry Breaking in the suppressed  Supergravity Theory:}    Given that we have chosen the potential and flipped the Higgs multiplets appropriately, so that $V$ takes the form (\ref{complexV3}), we now have the following equations for the VEVs:
\be
  \lt < \fr{\pa W}{\pa z^j} \rt > =0
\la{veveqforsuppsusy}
  \ee
This is just like the situation for the rigid supersymmetric theory, except that now the scalar potential is coupled to gravity.
In particular, these equations imply that the cosmological constant is exactly zero after the gauge symmetry breaking here. Of course, we cannot expect that this will remain the case after one loop, because there is no supersymmetry to cause cancellation of bubbles with each other here.

  \refstepcounter{orange}
{ \theorange}.\;
\la{GGFparagraph}
{\bf Gauge fixing and the \ME:}  In order to define the path integral, it is necessary to fix the gauges of the three gauge particles (vectors, gravitino, graviton) in these supergravity theories.    The procedure for vector boson gauge fixing  is well known.  We assume that gauge symmetry is not spontaneously broken in the original theory, or in the new suppressed  theory, at first.  After we have written down the new suppressed  SUSY action, we can examine the spontaneous breaking of gauge symmetry in that theory, and then we can derive the form for the Gauge Fixing and Ghost (`GGF') action using an auxiliary field $Z^a$ ( $U^a$ below is a Zinn source for the variation of the antighost $\h^a$, and $\a$ is a gauge parameter.):

\[
{\cA}_{\rm GGF\;From\;Z} =  
- \fr{1}{2\a} \int d^4 x  \lt (\pa^{\m} V_{\m}^{ a}+  \a G^a+ U^a\rt )^2 
 -\int d^4 x \]\be
  \lt \{ \h^a \pa^{\m} \lt ( \cD_{\m}^{ab} \w^b 
+ \lam^{a \a} \s_{\m \a \dot \b} \oC^{\dot \b}
+ C^{\a} \s_{\m \a \dot \b} {\ov \lam}^{a \dot \b}
+  \a \d G^a \rt )  
\rt \} 
\la{GGFaction} 
\ee
where the `would--be Goldstone Bosons  $G^a$ are ($m v^j$ is the VEV of the scalar field $A^j$): 
\be
 i g m T^{ai}_j \ov v_i A^j - i g m T^{ai}_j \A_i v^j = G^a
\ee
This makes no change to our conclusions above.  But the VEVs do depend on the flipping of the scalar fields.

\refstepcounter{orange}
{ \theorange}.\;
{\bf The Scalar Potential and the {\gust}:}  
Note the interesting form of (\ref{formofhiggssuperpot4}).  It is quite well suited to generating interesting solutions for (\ref{veveqforsuppsusy}).  The consequences of this, and the other subjects of this paper,  have been explored a little in
\cite{Dixon:2018dzx,Dixon:2017eej,Dixon:2019vhb}. Those papers were practical examples to prepare for the present more abstract treatment.

 \refstepcounter{orange}
{ \theorange}.\;
{\bf  Philosophy  of \sus:}  
The popular view, after 40 years of SUSY, is that SUSY implies supermultiplets unless SUSY is spontaneously or explicitly broken \ci{xerxes,haber,buchmueller}.
 \sS\ says there is another, easier, route.
In fact, \sus\ really says that the four problems are better viewed as four hints about how to analyze  supersymmetry in quantum field theory, given the experimental results. 
   
  \refstepcounter{orange}
{ \theorange}.\;
\la{parconclus}
{ \bf Conclusion and Summary:}
So in the end, \sus\ tells us how to convert a \gusgr\ into a theory that looks like an ordinary \gut, but that \gut\ is very special. We call it a suppressed  SUSY theory. 
 Here is a summary of the progress:
\ben
\item
The basic operation of  \sus\ is to `dequantize' a scalar or spin $\fr{1}{2}$ field, while simultaneously `quantizing' the conjugate Zinn source (which is the source for the BRS variation of that field, as used in the action that generates the Master Equation).
Any operation of this kind leaves the action and the \ME\ invariant--because it does not change the action at all, and then the path integral can be used to rederive the \ME, as explained in Appendix \ref{flippinginv}.
\item
In order to end up with the nice simple scalar potential, the original \gusgr\ must have a very special superpotential in the Higgs sector. It must have the form
 \be
  W_H =
 e^{\fr{-1}{4}\k^2  I_2 } 
M_P^{\fr{3}{2}}  \sqrt{I_3 }  
\la{formofhiggssuperpot4again}
    \ee
where $I_2$ is the quadratic invariant expression in the Higgs fields, and  $I_3$ is a general cubic invariant expression in the Higgs fields.
In addition, the \sus\ of the Higgs sector must reduce the quantized Higgs fields to a set so that the Higgs superpotential is real, as discussed in Paragraph \ref{quadsumforgravpot}.  
\item
The   new Feynman rules that arise for the new suppressed 
  \gusgr\ still yield a unitary theory, because the new quantized antighosts defined by  \sus\ do not propagate, and there is still a \ME\ that governs the theory.  
Thus  the new theory acts, to some extent, as though the suppressed fields and Zinn sources had simply been dropped, except that they are needed to work with the remaining fields so that the \ME\ stays true.   
Once we have generated the new suppressed \gusgr, we want to change notation to reflect which fields are still quantized, and then shift to a new vacuum, with all the consequences of that. 
\item
 The resulting, very constrained, physical theory can be obtained by setting all the Zinn sources to zero, leaving just the quantized fields, ghosts and antighosts.The resulting theory  does not have supermultiplets, so spontaneous breaking of SUSY is not needed.  It has a zero cosmological constant at tree level together with lots of spontaneous gauge symmetry breaking. It also still has a gravitino, which is now massive. Replacing the Zinn source terms restores the \ME, but of course the Feynman rules are different from the original theory.
\item
The \ME\ survives flipping, and it gives rise to a nilpotent cohomology operator (the `square root' of the \ME) that governs the perturbation expansion, in particular at one loop, as in \ci{olddixie}. This needs analysis, but that is not an easy task for this complicated operator. 
\item
The one loop corrections are a mystery at present and no explanation is offered for the hierarchy problems. But we should remember that conventional SUSY has the four problems already at tree level. So the suppressed GUSTs are progress, but the derivation of these new theories from the superstring is a major outstanding issue.
\een

 \begin{center}
{
Acknowledgments\\[.5cm]} 
\end{center}
It is a pleasure to thank Carlo Becchi, William Deans, Pierre Ramond, Peter Scharbach, Xerxes Tata and John C. Taylor 
for useful correspondence and conversations.

  \begin{center}
{ 
Appendices\\[.5cm]} 
\end{center}

   \refstepcounter{apple}
{ Appendix \theapple}.\;
\la{Ftermsinteg}
{\bf Quadratic Terms from the Rigid SUSY Action:}
We will illustrate this for one example: the F auxiliary terms.  Start with the action for rigid SUSY. This includes the following terms:
\be
\int d^4 x \lt \{ 
 F^i {\ov F}_i + Y_i^{ \a} F^i C_{\a} + {\ov Y}^{i \dot \a} \oF_i \oC_{\dot \a}
+ F^i \fr{\pa W}{\pa A^i}
+{\ov F}_i \fr{\pa {\ov W}}{\pa \A_i} \rt \} 
\la{Ftermsinaction}
\ee
These are the terms involving the auxiliary F.  They come from the superpotential terms and also the terms $Y_i^{ \a} \d \chi_{i \a}=Y_i^{ \a}\lt ( F^i C_{\a} + 
  T^i_{aj} \chi^i_{\a}\w^a 
+ \cD^i_{j\a\dot \b} A^j\oC^{\dot \b}\rt ) 
$, which are the usual BRS SUSY variation in the Wess Zumino gauge for rigid SUSY, coupled to gauge theory. We put this into the first line of (\ref{origpat}) and then complete the square
as follows:
\be
\int d^4 x \lt \{ 
\lt ( {\ov F}_i + Y_i^{ \a} C_{\a} + \fr{\pa W}{\pa A^i} \rt )
\lt (F^i + {\ov Y}^{i \dot \a}  \oC_{\dot \a}
+ \fr{\pa {\ov W}}{\pa \A_i} \rt )
\la{firstlineoffcompletion}
\ebp-
\lt ( Y_i^{ \a} C_{\a} + \fr{\pa W}{\pa A^i} \rt )
\lt (  {\ov Y}^{i \dot \a}  \oC_{\dot \a}
+ \fr{\pa {\ov W}}{\pa \A_i} \rt )
\rt \} 
\la{secondlineoffcompletion}
\ee
Shifting the variable $F^i$ in line (\ref{firstlineoffcompletion}) to get it to the form $F^i \ov F_i$,  and then integrating the path integral  in the first line of (\ref{origpat}) over $F$ and $\ov F$, yields a constant, which we can ignore in the path integral.  The second term (\ref{secondlineoffcompletion}) remains and we use it as the term (\ref{AfromF}). This second term also makes its appearance in the expression 
$\cA[{\rm Q}, \cZ_{\rm Q} ] $ in the second line of (\ref{origpat}).  These terms act as if they had the variations of the auxiliaries\footnote{In fact, it is possible to introduce such terms as these even when there are no auxiliaries known, and still get a \ME\ \ci{Dixon:1991wt}.}, which is why the identity (\ref{mesimplezero}) holds. Other auxiliary terms, such as $D^a$ and $Z^a$ work in exactly the same way, yielding the terms (\ref{AfromD}) in paragraph \ref{theAterms}, and (\ref{GGFaction}) in paragraph \ref{GGFparagraph}.   
Supergravity has additional auxiliaries that need integration \ci{freepro}, and they work the same way.

      \refstepcounter{apple}
{ Appendix \theapple}.\;
\la{brstme}
{\bf Formal Derivation of The BRST \ME:}
To derive this, one must start with a nilpotent set of BRST transformations, labelled $\d$,  for the fields, ghosts, and antighosts $ {\rm Q}= \cF, G,\cP$,  and any auxiliary fields ${\rm A_F}$. See paragraph  \ref{mastereqpara1} for the definition of these symbols.  We assume that this is possible for the  GUST of interest  \ci{babble}.  Then one can simply take the path integral:
\[
Z =  \P_{Q(x),A_F(x) } d {\rm Q}(x)\;  d    {\rm A_F}(x)\; e^{\fr{i}{\hbar} \lt \{ \cA_{\rm Q} + \int d^4 x \;\cZ_{\rm Q} \d \rm Q+  \int d^4 x \;S_{\rm Q} \rm Q \rt \}}
\]\be
=e^{\fr{i}{\hbar} \cW[\cZ_{\rm Q} ,S_Q]}=\P_{Q(x)} d {\rm Q}(x) \; e^{\fr{i}{\hbar} \lt \{ \cA[{\rm Q}, \cZ_{\rm Q} ] +  \int d^4 x \;S_{\rm Q} \rm Q \rt \}}
\la{origpat}\ee
where sums and products over indices and names of variables are suppressed. 
 $\cA_{\rm Q}$ is an invariant action which is a function of the quantized variables $Q$ and the auxiliaries ${\rm A_F}$. 
  $\cZ_{\rm Q} $ are Zinn sources   which are 
multiplied by the variations $\d {\rm Q}$ of the quantized variables.  
  $ S_{\rm Q} $ are  ordinary sources which are multiplied by the quantized variables ${\rm Q}$. Some simple sign juggling is needed for anticommutation in appropriate places.
We can   perform the integrations over  ${\rm A_F}$ easily, as in Appendix \ref{Ftermsinteg}. This leads to quadratic terms in the action like equations (\ref{AfromF})
and (\ref{AfromD}) in paragraph \ref{theAterms}, and (\ref{GGFaction}) in paragraph \ref{GGFparagraph}.    We write $\cA[{\rm Q}, \cZ_{\rm Q} ] $ for the result of integrating the auxiliaries. This is always a local action.
From the nilpotence and the invariance we can immediately see that a simple shift of the integration variables by $Q \ra Q+\varepsilon \d Q, A_F \ra A_F+\varepsilon \d A_F$, 
(the constant anticommuting $\varepsilon$ just preserves the Grassmann character) yields an identity: 
\be
\d \cA_{\rm Q} =  \d^2 Q =0
\Ra  
\int d^4 x  S_{\rm Q} \varepsilon \d Q=\varepsilon \int d^4 x  S_{\rm Q} \fr{\d \cW}{\d  \cZ_Q }
=0
\la{msterfstr}
\ee
Then we define the Legendre transforms  in the usual way \ci{zinnbook}:
\be
\cG = {\cal W} + \int d^4 x  S_{\rm Q}  Q
;\; \fr{\d \cG}{\d Q} =  S_{\rm Q} 
; \; \fr{\d \cW}{\d  S_{\rm Q}} =  -{\rm Q}
; \;\fr{\d \cG}{\d \cZ_Q} =   \fr{\d \cW}{\d \cZ_Q}
\ee
and then from (\ref{msterfstr}), we get the  BRST \ME, which is the shortened form of equation (\ref{brstmas}):
\be
\int d^4 x  \fr{\d \cG}{\d Q} \fr{\d \cG}{\d \cZ_Q}=0
\la{mesimple}
\ee

\refstepcounter{apple}
{ Appendix \theapple}.\;
\la{flippinginv}
{\bf Flipping Invariance of the BRST Master Equation:}
 We 
note that $\cG= \cA[{\rm Q}, \cZ_{\rm Q} ] + \hbar \cG_1 + \hbar^2 \cG_2 + \cdots $ where $\cA[{\rm Q}, \cZ_{\rm Q} ]$ is the result of integrating the auxiliary fields in (\ref{origpat}).   So in particular, from  (\ref{mesimple}): 
\be
\int d^4 x  \fr{\d \cA[{\rm Q}, \cZ_{\rm Q} ] }{\d Q} \fr{\d \cA[{\rm Q}, \cZ_{\rm Q} ] }{\d \cZ_Q}=0
\la{mesimplezero}
\ee
This can be viewed as an invariance in many different ways. For example 
we can define
\be
\d_1 Q = \fr{\d \cA[{\rm Q}, \cZ_{\rm Q} ] }{\d \cZ_Q}
\Ra 
 \d_1   \cA[{\rm Q}, \cZ_{\rm Q} ]=0
\la{delta1new} \ee
 or we could define 
\be
\d_2  \cZ_{\rm Q}  = \fr{\d \cA[{\rm Q}, \cZ_{\rm Q} ] }{\d Q}
\Ra 
 \d_2   \cA[{\rm Q}, \cZ_{\rm Q} ]=0
 \ee
 We could also choose any other combination of  $Q$ and $\cZ_{\rm Q}$ as the variables for this new definition of $\d$, provided we take only one variable from each conjugate pair.    The invariance of this action is indifferent to our choice of what the quantized variables are, and what the Zinn sources are.  So we can flip between them any way we want, and still get a derivation like the one in Appendix 1, provided we integrate over the appropriate new quantized variables, and vary the source terms in (\ref{origpat}) for the quantized variables from  $\int d^4 x S_Q Q$ to the set of the new quantized variables\footnote{Note, for example, that we could rederive (\ref{mesimple}) by using the variations (\ref{delta1new}). These variations arise after the auxiliaries are integrated.}.  In each case we get a different series $\cG= \cA[{\rm Q}, \cZ_{\rm Q} ] + \hbar \cG_1 + \hbar^2 \cG_2 + \cdots $, because the terms other than the action $\cA[{\rm Q}, \cZ_{\rm Q} ]$ depend on the Feynman rules, which depend on which variables are quantized.  But the form of the BRST \ME, and the form of the action  $ \cA[{\rm Q}, \cZ_{\rm Q} ]$ are the same for all of these choices, even though the series $\cG$ differs for each new choice, because the quantized variables differ.  This derivation needs more  attention to the Jacobians \ci{Dixon:2018dzx}.

      \refstepcounter{apple}
{ Appendix \theapple}.\;
\la{propdestroysunitarity}
{\bf Suppression keeps unitarity only for scalar and spinor flips.}
  As discussed often in the main article here, if a flipped physical field variable $\cF$ has an antighost propagator involving $\cZ_{\cF}$, then there are new non-trivial Feynman diagrams with the propagator for the relevant antighost. This is an issue for  vectors, gravitinos or gravitons, but not for scalars and spinors. When there are new antighost propagators, there are no unphysical fields to be cancelled in the usual Faddeev-Popov sense, since the conjugate physical fields have been converted to  unphysical Zinn sources.  That means we must lose unitarity for such cases.  That is why suppression is useful only  for scalars and spinors. Note that this is also  a problem for gauge fixing antighosts and their Zinn sources as in equation (\ref{GGFaction}), where there is clearly a new $U U  $ propagator when $U$ gets quantized, mixing with other fields.  The case of flipping ghosts might need some attention.

      \refstepcounter{apple}
{ Appendix \theapple}.\;
\la{propforlambdaappendix}
{\bf Does the antighost from the flipped Zinn Source $\Lambda$ Propagate?}
To try to analyze this in general is not practical, because the issues have to do with multiple irreducible representations.  Let us look at a specific example from 
\cite{Dixon:2018dzx, Dixon:2017eej, Dixon:2019vhb}. We start with the following conjugate pairs of scalars and Zinn sources in the original Zinn action before we do any flipping:
\be
\Lambda_{L}^i 
\d H_{Ri}  
+
\Lambda_{Ri} 
\d H_{L}^i 
+
\Lambda_{S}^a 
\d S^a 
+ {\rm c.c.}\; {\rm where}\; i = 1 \cdots 5,\; a = 1 \cdots 24\ee
In the above, $H_{L}^i$ is a 5, $H_{Ri}$ is a $\ov 5$ and $S^a$ is a complex 24 of SU(5). Their conjugate Zinn sources $\Lambda$ have quantum numbers to match.   
Then we look to see whether the term of the kind in Paragraph \ref{propforlambdaparagraph} appears or not, after flipping, and after a VEV is chosen in the suppressed GUST. 
This is easiest to see if we introduce new variables to see what is quantized and what is not in the suppressed GUST.  After flipping this becomes
\be
\Lambda_H^i 
\d H_{i}  
+
\eta_{H i} 
\d J_H^i 
+
\Lambda_{K}^a 
\d K^a 
+
\h_{K}^a 
\d J_K^a 
+ {\rm c.c.} \ee
where the $\eta_H,\eta_K$ are new antighosts, the
 $\Lambda_H,\Lambda_K$ are new Zinn sources,   the $H,K$ are new quantized fields  and the $J_H,J_K$ are new Zinn sources.  The rotations in the new suppressed  BRST variations ($\d$) rotate new quantized fields into new quantized fields, and they rotate new Zinn sources into new Zinn sources.  
So the relevant terms are
\be
\Lambda_H^j 
iT^{ai}_j H_{i}  \w^a 
+
\eta_{H i} 
i T^{ai}_j J_H^{j} \w^a  
+
\Lambda_{K}^a 
f^{abc} K^{b} \w^c  
+
\h_{K}^a 
f^{abc}  J_S^{b} \w^c  
+ {\rm c.c.} 
\la{newzinnandantighostterms}
\ee
If there were any propagator type terms of the type (\ref{chiralzinnxitermsvev})  here, they would have to come from 
the $
\eta_{H i} 
i T^{ai}_j J_H^{j} \w^a  
+
\h_{K}^a 
f^{abc}  J_S^{b} \w^c  
+ {\rm c.c.} $ terms in (\ref{newzinnandantighostterms}).  But of course $J_H^{j}$ and $J_S^{b}$ do not get VEVs, because they are new Zinn sources in the suppressed theory. VEVs do appear in the new fields $H$ and $K$, but those are coupled to the new Zinn sources  $J_H^{j} $ and $J_S^{a} $ in  (\ref{newzinnandantighostterms}), so they do not give rise to propagators for the new antighosts either. So we see that for this particular model, the new antighosts $\eta_{H i}, 
\h_{K}^a $ do not propagate, even after the generation of the new VEVs.

{\tiny (Version:
\articlenumber, \today,  
\hourandminute)}


\begin{thebibliography}{99}
\bibitem{babble} We want the \ME\ of an \gust\ or an \gutt, or others of that kind. No such \ME\ has ever been explicitly written down in full form.  There are unsolved issues  with   gravitino and graviton gauge fixing, but we assume they cause no problems here.
\bibitem{Becchi} I find this general and useful remark easy to grasp and so I like it.  It was made a few years ago by Carlo Becchi, prompted by a request from John C. Taylor for a general statement. I have the impression that C.  Becchi prefers his more circumspect analysis in \ci{becchiarticles2}.  
It is clear that  the Master Equation (Becchi calls it the Zinn identity) 
has many features that are powerful and subtle, as does the Poisson Bracket, which it closely resembles.
\end{thebibliography}
 \end{document}